\documentclass[preprint,12pt]{elsarticle}
\usepackage{graphicx}
\usepackage{amssymb,amsmath}
\usepackage{color}
\usepackage{enumitem}
\usepackage{gensymb}
\usepackage{subcaption} 
\usepackage{siunitx}
\usepackage[colorlinks=true, allcolors=black]{hyperref}

\usepackage{booktabs}

\usepackage{xcolor}     
\begin{document}
\begin{frontmatter}
\title{Graph Neural Networks to Predict Coercivity of Hard Magnetic Microstructures}


\author[inst1]{H. Moustafa\corref{c1}}
\cortext[c1]{Corresponding author}
\ead{heisam.moustafa@donau-uni.ac.at}
\author[inst1]{A. Kovacs}
\author[inst1]{J. Fischbacher}
\author[inst1]{M. Gusenbauer}
\author[inst1,inst2]{Q. Ali}
\author[inst1]{L. Breth}
\author[inst1,inst2]{T. Schrefl}
\author[inst1]{H. Oezelt}

\address[inst1]{
  Department for Integrated Sensor Systems, University for Continuing Education Krems,
  Viktor Kaplan-Straße 2, Wiener Neustadt, 2700, Austria
}

\address[inst2]{
    Christian Doppler Laboratory for Magnet Design through Physics-Informed Machine Learning, 
    Viktor Kaplan-Straße 2, Wiener Neustadt, 2700, Austria
}

\begin{abstract}
 Graph neural networks (GNN) are a promising tool to predict magnetic properties of large multi-grain structures, which can speed up the search for rare-earth free permanent magnets. In this paper, we use our magnetic simulation data to train a GNN to predict coercivity of hard magnetic microstructures. We evaluate the performance of the trained GNN and quantify its uncertainty. Subsequently, we reuse the GNN architecture for predicting the maximum energy product. Out-of-distribution predictions of coercivity are also performed, following feature engineering based on the observed dependence of coercivity on system size.
\end{abstract}

\end{frontmatter}


\section{Introduction}
With the rapid progress of climate change, it is more urgent than ever to accelerate the \textit{green transition}, especially in energy production e.g. offshore wind turbines and transportation e.g. electric vehicles. Consequently, the demand for the rare-earth element neodymium, main component in NdFeB permanent magnets, will rise more than threefold until 2050 
Since rare-earth elements are critical regarding price and supply stability, the content of rare-earth elements in such magnets needs to be reduced while maintaining performance. Micro- and nanostructural design was identified as the most promising strategy to achieve this goal \cite{skomski_magnetic_2016} \cite{poudyal_advances_2013}. To examine the effect of microstructural design changes on the magnet's performance, simulations are a good means. To reduce the computational effort of simulations, in this work we focus on a graph neural network (GNN) layout to do an evaluation of the influence of different microstructural parameters on the magnet's performance. GNNs have been shown to offer a favorable combination of architectural simplicity, predictive accuracy, and strong generalization capabilities when simulating complex physical systems  \cite{sanchez-gonzalez_a_godwin_j_pfaff_t_ying_r_leskovec_j_battaglia_pw_learning_2020}.

Dai et al. \cite{dai_graph_2023} created a graph neural network for predicting polycrystalline material properties, namely all three diagonal components of the effective ion conductivity matrix. The network was trained and tested on a dataset of over 5,000 microstructures with finite-width grain boundaries, generated by Voronoi Tessellation and electron backscatter diffraction image simulation. The optimized model outperformed a linear regression model and two convolutional neural networks. They showed the feature importance of grain and edge features, investigated the model's generalization capabilities, and transfer learning performance.

GrainNN is a GNN designed by Qin et al. \cite{qin_grainnn_2023} to predict the evolution of grain microstructures in metals during additive manufacturing. It is implemented with an LSTM-based regressor and classifier. The developed GNN generalizes to domain size, number of grains, and initial grain parameters while being significantly faster than phase field simulations.

PolyGRAPH presented by Hestroffer et al. \cite{hestroffer_graph_2023} is a GNN to predict yield strength and stiffness of polycrystalline materials. The dataset consisted of 1200 microstructures. As node features, only the crystallographic orientation and grain size were used, no edge features were implemented. The architecture consists of one fully connected layer for feature pre-processing, two message-passing layers using the GraphSAGE algorithm \cite{hamilton_inductive_2017}, a global mean pooling layer, and two fully connected post-processing layers. A good prediction accuracy was achieved.

\begin{figure}[htb]
    \centering
    \includegraphics[width=0.8\columnwidth]{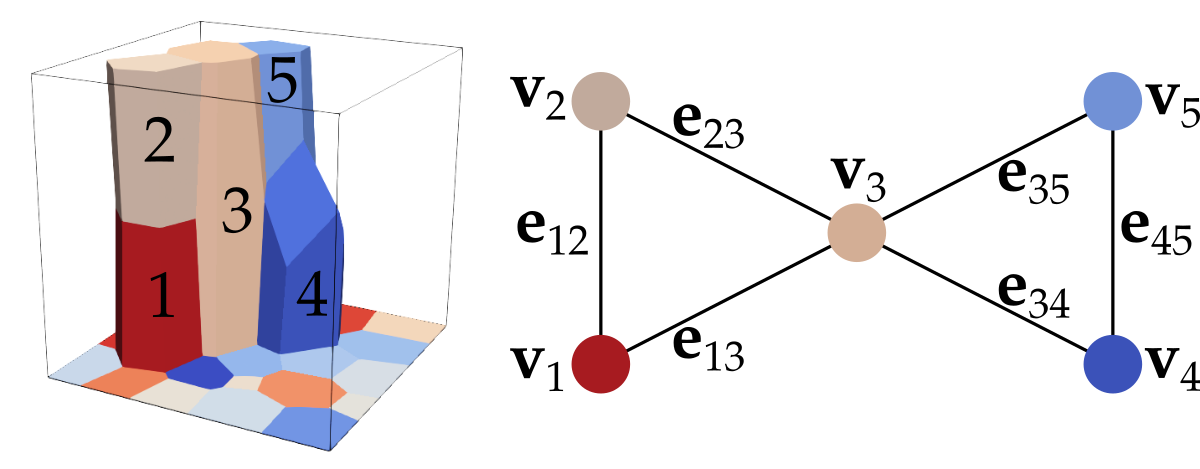}
	\caption{Left: five selected grains of a cuboid grain structure. Right: The nodes correspond to the grains, including the grain features $\mathbf{v}_i$. The edges $\mathbf{e}_{ij}$ correspond to the grain boundary phases between each neighbouring grain. 
 }
	\label{GNN example}
\end{figure}

While GNNs have been successfully applied to predict various properties of polycrystalline materials, their use for coercivity prediction has not yet been investigated.
We aim to predict the coercivity of hard magnetic Nd\textsubscript{2}Fe\textsubscript{14}B microstructures using a graph neural network (GNN) with edge features, trained on data generated through simulations. The architecture is implemented using the PyTorch framework \cite{NEURIPS2019_9015} along with the PyTorch Geometric library \cite{fey_fast_2019}. We suggest a GNN to be beneficial since it captures the neighbourhood structures in the microstructure. The message passing between neighbors of a GNN enables incorporating physical interactions among the grains and grain boundaries. Fig.~\ref{GNN example} shows the physical model of a magnetic microstructure and its representation as a graph.

The paper is structured as follows. Section~\ref{sec: training data generation} describes the dataset generation for training, validation, and testing using Neper \cite{quey_large-scale_2011}, an open source software for polycrystal generation, the open source platform Salome for meshing \cite{salome}, and a reduced order micromagnetic simulation software \cite{moustafa_reduced_2024}. The section also provides information on the key characteristics and structure of the resulting dataset. Section~\ref{sec: data preprocessing} describes data preprocessing methods for machine learning. The dataset is scaled, and a correlation matrix is generated to identify strongly correlated features. Permutation feature importance helps to identify the features that are important to the trained model. Section~\ref{sec: graph neural network design} gives a brief overview of the GNNs design. Furthermore, it describes the used layers with activation functions, the backpropagation mechanism, and the hyperparameter training in more detail. Finally, hyperparameter tuning is described. In section~\ref{sec: uq with bnn} we present the theoretical basis of uncertainty quantification (UQ) with the help of a Bayesian GNN. These concepts are then applied to our specific GNN architecture, and the resulting uncertainties are discussed. In section~\ref{sec: results} the results for the coercivity prediction, the maximum energy product prediction, and the out-of-distribution generalization are presented and discussed. An investigation of the influence of the microstructure's edge length on the coercivity is done in section~\ref{sec:size}. The conclusion is drawn in section~\ref{sec: conclusion}.

\section{Training Data Generation}
\label{sec: training data generation}
Training data for the GNN are generated by our reduced order model (ROM). The ROM significantly reduces computational demands by using a 2D surface mesh, approximating grain reversal with analytic computations of the local Stoner-Wohlfarth switching field, and analytically calculating the demagnetization field, using formulas for polyhedral geometries \cite{singh_new_2001}:
\begin{equation}
     \mathbf{h}_\mathrm{demag}(\mathbf{x}) = \frac{1}{4\pi}\sum_j\sum_k\int_{S_{jk}}\mathbf{n}_{jk}(\mathbf{x'})\cdot\mathbf{m}_j\frac{\mathbf{x}-\mathbf{x'}}{|\mathbf{x}-\mathbf{x'}|^3}dS'.
\end{equation}
The magnetostatic field $\mathbf{h}_\mathrm{demag}$ at point \textbf{x} is the sum of surface integrals over all polyhedral surfaces $S_{jk}$. The index $j$ runs over all grains and $k$ over the surfaces of grain $j$. Vector $\mathbf{n}_{jk}$ is the outer normal and $\mathbf{m}_j$ the magnetization. With the use of hierarchical matrices \cite{mikhalev_iterative_2016} computational effort is further reduced. An assumption that is made when using the reduced order model is that it assumes hard magnetic defect-free grains with a non-magnetic grain boundary.

\begin{figure}[htb]
    \centering
    \includegraphics[width=1\columnwidth]{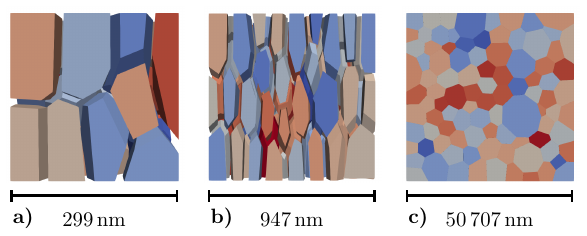}
	\caption{Three different microstructures of the created dataset. a) grain boundary thickness: \SI{9}{\nano\metre}, aspect ratio: 1.4, b) grain boundary thickness: \SI{30}{\nano\metre}, aspect ratio: 3.0, c) grain boundary thickness: \SI{8}{\nano\metre}, aspect ratio: 1.0.
 }
	\label{fig: three tesselations}
\end{figure}

Using the ROM, 1054 different cubic microstructures with an edge length varying between 179\,nm and 68326\,nm, tesselated into 2 to 3375 grains by the software Neper, were simulated. The microstructures differ in the grains' aspect ratio, which is defined as the width of the grain divided by its height, and in the thickness of the non-magnetic grain boundary phase. The grains' arrangement and magnetocrystalline easy axis direction within the cuboid differ as well. The easy axis of each grain is randomly oriented within a predefined maximum cone angle. To ensure a uniform distribution over the cone, we use the method proposed by Marsaglia \cite{marsaglia_choosing_1972}. For all grains, the material is Nd\textsubscript{2}Fe\textsubscript{14}B with the intrinsic magnetic properties at room temperature: spontaneous polarisation \SI{1.6}{\tesla}, magnetocrystalline anisotropy constant \SI{4.9e6}{\joule\per\metre\cubed}, exchange stiffness constant \SI{8e-12}{\joule\per\metre}. Fig.~\ref{fig: three tesselations} depicts three microstructures of the created dataset.

\begin{table}[ht]
\centering
\caption{The feature ranges.}
\begin{tabular}{lccc}
    \textbf{Node Feature} & \textbf{Minimum} & \textbf{Maximum} & \textbf{Mean} \\ 
    \midrule
    {Coordinate x in \SI{}{\nano\metre}}& 5.7 & 67085 & 3767 \\
    {Coordinate y in \SI{}{\nano\metre}}& 5.2 & 67109 & 3754 \\
    {Coordinate z in \SI{}{\nano\metre}}& 7.4 & 64224 & 3790 \\
    {Sphericity} & 0.32 & 0.95 & 0.78 \\ 
    {Area in \SI{}{\nano\metre\squared}} & \SI{3.4e3}{} &  \SI{3.4e9}{} &  \SI{28e6}{} \\ 
    {Avg. dihedralangle in \SI{}{\degree}} & 70 & 145 & 113 \\ 
    {Aspect ratio} & 0.11 & 38.75 & 2.58 \\ 
    {Edge length in \SI{}{\nano\metre}} & 179 & 68326 & 7532 \\
    {Easy theta in \SI{}{\radian}} & 0  &  1.39 & 0.23 \\
    {Easy phi in \SI{}{\radian}} & $-\pi$  &  $\pi$ & 0 \\
    {Distance from centre in \SI{}{\nano\metre}}&0 &51938 &3504 \\
    {SW switching field in \SI{}{\ampere\per\metre} }& \SI{3.0e6}{} & \SI{6.0e6}{} & \SI{3.7e6}{} \\ 
    {Number of Grains} & 2 & 3375&184\\
    \bottomrule
    \textbf{Edge Feature} & \textbf{Minimum} & \textbf{Maximum} & \textbf{Mean} \\ 
    \midrule
    {Grain boundary thickness in \SI{}{\nano\metre}} & 1 & 30 & 11.5 \\
    \bottomrule
\end{tabular}
\label{tab:features}
\end{table}

The computed dataset is split into a training set, a validation set, used to validate after each epoch, and a test set, used once the training is done and the weights and biases are set for the model to validate the trained model with unseen data. The sets contain 734 (70\,\%), 157 (15\,\%), and 158 (15\,\%) graphs, respectively. 

\section{Data Preprocessing}
\label{sec: data preprocessing}
\begin{figure}
    \centering
	\includegraphics[width=0.5\textwidth]{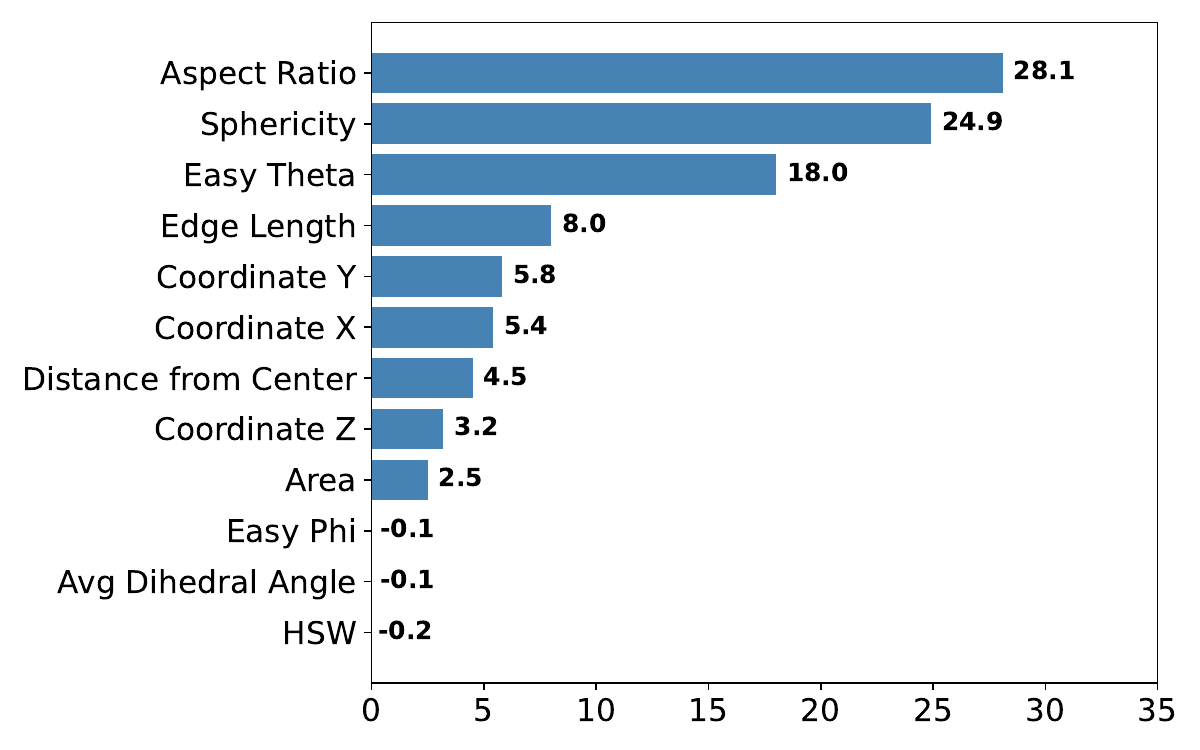}
	\caption{The relative permutation feature importance of the grain features; the sum of the importances is 100.}
	\label{fig: feature importance}
\end{figure}

Preprocessing of the data mostly enhances the performance of the GNNs. One important preprocessing step is feature scaling. Machine Learning algorithms don't perform well when the input features have very different scales \cite{geron_hands-machine_2019}. This is because features with larger values would dominate the training process, since the gradient calculation incorporates the feature value. This means that the optimizer focuses on the larger-scale features, updating its weights and ignoring smaller-scale features. Standard scaling of feature vector $\mathbf{x}$ is performed as:
\begin{equation}
    \mathbf{x_{sca}} = \frac{\mathbf{x} - \mu}{\sigma}
\end{equation}
where $\mu$ and $\sigma$ are the mean and the standard deviation of feature vector $\mathbf{x}$, respectively. This transformation is applied to the node features of our GNN, using the scikit-learn's StandardScaler \cite{scikit-learn}. $\mathbf{x_{sca}}$ represents the normalized form of the feature vector, adjusted to have a mean of 0 and a standard deviation of 1. The scaler is first fitted using only the training data and subsequently applied to scale the entire dataset including the validation and test set. This procedure ensures that there is no scaling bias from the training set to the validation or test set. Standard scaling is applied to the labels as well, MinMaxScaling, to avoid negative values, to the edge features. These scaling techniques were chosen based on empirical observation.\\
A correlation matrix is used to assess the relationships between input features, allowing for the identification of highly correlated features. The Pearson correlation coefficient $\rho_{x,y}$ of two features (x,y) is defined as:
\begin{equation}
    \rho_{x,y} = \frac{\mathrm{cov}(X,Y)}{\sigma_x\sigma_y}
    \label{eq: pearson}
\end{equation}
and used as a correlation metric. $\mathrm{cov}$ is the covariance and $\sigma$ is the standard deviation of the respective feature. The range of the Pearson coefficient is from -1 to 1. A value near 1 implies that there is a strong positive correlation, a value of zero implies no linear dependency, and a value near -1 implies a strong negative correlation. Features with an absolute correlation coefficient over 0.8 have such a strong correlation that one of them can be removed from the dataset.
Since features such as equivalent diameter, area, and volume are inherently correlated, the volume and equivalent diameter were excluded. Further empirical observation led to the use of the features named in Tab.~\ref{tab:features}. An explanation of selected features is given below to aid interpretation:
\begin{itemize}
    \item Coordinate X, Y, and Z: the center positions of individual grains.
    \item Sphericity: the ratio of the surface area of a sphere with an equivalent volume to the surface area of the grain.
    \item Area: the grain surface area.
    \item Dihedral angle: the angle between two adjacent surfaces of a grain.
    \item Aspect ratio: the ratio of grain length to grain width.
    \item Edge Length: the edge length of the cubic microstructure.
    \item Easy theta and Easy phi: the respective polar and azimuthal angle components of the easy axis.
    \item Distance from Center: the distance between the grain center and the microstructure center.
    \item SW switching field: the Stoner-Wohlfarth switching field of the respective grain.
    \item Number of Grains: The number of grains per microstructure.
\end{itemize}
The features were selected by gradually removing them and checking the impact on the model's prediction accuracy on the validation set.\\
After training a first model, a permutation feature importance analysis is done to measure the contribution of each feature to a fitted model's performance. The values of one feature are randomly shuffled and the resulting model degradation is observed. If the degradation of the model is high, the feature importance is high as well. It is important to note that features which have a low importance for a bad model can have a high importance for a good model. Feature importance does not reflect the intrinsic predictive value of a feature but the importance for the particular model. Permutation feature importance can be calculated on the training set or on a held-out set. Features that demonstrate significance during training but lack importance on a held-out dataset can lead to overfitting, reducing the model's ability to generalize effectively. \cite{noauthor_42_nodate} Fig.~\ref{fig: feature importance} is depicting the computed feature importance for the GNN. The whole contribution of the features to the model is 100. Aspect ratio and sphericity have the highest importance according to the analysis. This is presumably attributed to the strong impact of the shape on the magnetostatic field, which in turn plays a critical role in determining coercivity. The easy axis azimuthal phi angle, which is the angle of rotation of the radial line around the polar axis theta, the average dihedral angle, and the Stoner-Wohlfarth switching field exhibit a nearly negligible influence on the trained model.

\section{Graph Neural Network Design}
\label{sec: graph neural network design}
In this section, we present an overview of the GNN architecture. We describe the individual layers used in subsection~\ref{ssec:layers}, explain the backpropagation mechanism in subsection~\ref{ssec:backpropagation}, and conclude in subsection~\ref{ssec:hyper} by addressing hyperparameter tuning.

The GNN consists of \textbf{(a)} 1x embedding layer, where the 12 selected node features are embedded in an 18-dimensional space through a linear transformation. This ensures that complex relationships can be captured by the model. Next, \textbf{(b)} one convolutional layer which operates according to Eq.~\ref{eq:convlayer} makes sure that information from the neighbours is taken into account. One fully connected linear layer \textbf{(c)} linearly transforms the results, in the global mean pooling layer \textbf{(d)} each graph's node features are aggregated. To generate the final output, three fully connected layers \textbf{(e)} reduce the dimensionality.\\
The GNNs architecture overview:
\begin{enumerate}[label=(\alph*)]
\item 1x Embedding Layer
\item 1x Convolutional Layer
\item 1x Fully Connected Layer with Dropout
\item 1x Global Mean Pooling Layer
\item 3x Fully Connected Layer
\end{enumerate}

\begin{figure*}
	\includegraphics[width=\textwidth]{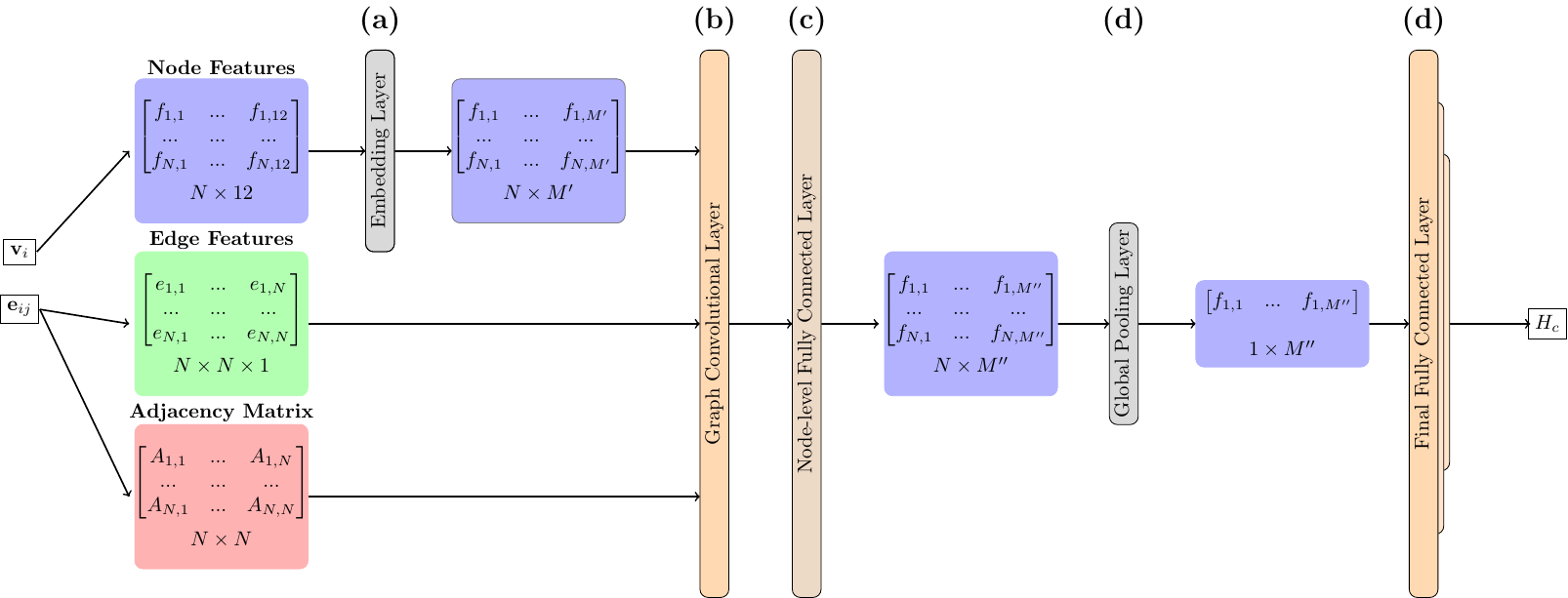}
	\caption{The GNN architecture inspired by \cite{dai_graph_2023}. $\mathbf{v}_i$ are the grain feature vectors of one graph, $\mathbf{e}_{ij}$ the corresponding edges and edge features. $N$ is the number of grains per graph, $M'=18$ is the embedded node feature length, $M''$ has the length 16. After the global mean pooling layer, only one node per graph remains. Three fully connected layers serve as the decoder for the coercivity prediction of the GNN.}
	\label{GNNArch}
\end{figure*}

\subsection{Layers}
\label{ssec:layers}
The \texttt{torch.nn.Linear} layer applies a linear transformation to the input features:
\begin{equation}
y = x\mathbf{W}^T+b
    \label{eq:linearlayer}
\end{equation}
where $x$ are the input features, $\mathbf{W}$ the weights matrix with the size output features $\times$ input features, which are trained, $b$ the bias and $y$ the output. Without a bias term, the model would be restricted to fitting outputs that pass through the origin, limiting its ability to represent the data accurately.\\
Activation functions introduce non-linearity to the GNN. In our implementation, the output of a linear layer is passed through either the Rectified Linear Unit (ReLU) in the final fully connected layers \textbf{(d)} or the SoftPlus activation function in the node-level fully connected layer \textbf{(c)}.
ReLU has the behaviour
\begin{equation}
\text{ReLU}(x) = 
\begin{cases}
x, & \text{if } x > 0 \\
0, & \text{if } x \leq 0
\end{cases}
\end{equation}
SoftPlus activation function is a smooth approximation to the ReLU activation function:
\begin{equation}
    \mathrm{Softplus}(x) = \frac{1}{\beta} \log\left(1+e^{x\beta}\right).
    \label{eq:softplus}
\end{equation}
Where $\beta$ is set to 1 and values above 20 for $\beta x$ fall back to a linear function $\mathrm{SoftPlus}(x) \approx x$. Both activation functions always have a positive output.\\
Hinton et al. \cite{hinton_improving_2012} describe dropout as a simple regularization technique that improves the generalisation ability of NN and prevents overfitting. The dropout probability $p$, with which some elements of the input tensors are randomly zeroed out, is usually set to 50\%, but depends on the specific case. Dropout is generally applied before the output of a hidden fully-connected layer. However, it can also be beneficial to apply it to convolutional layers \cite{lai_analysis_2017}. After tuning, we applied dropout with a probability $p=0.6$ to the post convolution layer of our GNN, in other words the drop out is included between the layers \textbf{(b)} and \textbf{(c)} in Fig. \ref{GNNArch}.\\
We designed a customized convolutional layer, that was inspired by Dai et al. \cite{dai_graph_2023}, which concatenates the source and destination node features with the edge attributes, applies a linear transformation, and uses a ReLU activation:
\begin{equation}
 x'_i = \frac{1}{\mathcal{N}(i)}\sum_{j\in \mathcal{N}(i)} \mathrm{ReLU}((x_i ^\frown x_j ^\frown e_{ij})\cdot W^T+b).
    \label{eq:convlayer}
\end{equation}
Where $x'_i$ are the updated node features, $x_j$ the node features of the neighbours, $\mathcal{N}(i)$ the set of neighbours of node $i$, and $e_{ij}$ the respective edge attributes. $^\frown$ symbolises concatenation. The number of convolutional layers corresponds to the number of hops in the neighbourhood. A one-hop neighbourhood means that only the next neighbours are considered, in a two-hop neighbourhood also the neighbours' neighbours are considered. The influence of neighboring nodes diminishes with each hop. With the first convolutional layer, each node's features are updated by aggregating its own features and those of its immediate neighbors. In the second convolutional layer, node features are updated again in the same manner, but this time, the features of its neighbors also include information from their neighbors, thereby incorporating a broader neighborhood while the influence of distant nodes diminishes progressively.

Now, one could find it appealing to use a very high number of convolutional layers, to capture a broad neighbourhood. However, despite an increasing computational complexity, it will lead to over-smoothing; after many convolutional layers, the nodes may end up indistinguishable, having similar feature vectors. For our GNN, empirical observations suggest that optimal model performance is obtained using a single convolutional layer.
\subsection{Backpropagation}
\label{ssec:backpropagation}
\begin{figure}
    \centering
	\includegraphics[width=0.8\textwidth]{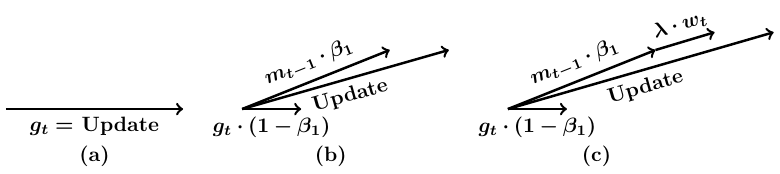}
	\caption{Weight updates for different optimizers: a) SGD, b) Adam, c) AdamW. The learning rate and -- for Adam and AdamW -- the denominator are omitted for simplicity. Here the gradient of the loss is denoted as $g_t = dL/dw_t$.}
	\label{fig: optimizers}
\end{figure}

Backpropagation is a crucial algorithm used in neural networks, it iteratively minimizes the loss function, e.g. MSE, by adjusting the weights and biases. These parameters are adapted in each epoch by following the error gradient. The weight update $\Delta w_{ij}$ for the weight from neuron $i$ to neuron $j$ is calculated using Eq.~\ref{eq:weightupdate}
\begin{equation}
\Delta w_{ij} = \eta \frac{dL}{dw_t} =  \eta \delta_j O_i 
\label{eq:weightupdate}
\end{equation}
where $dL/dw_t$ is the gradient of the loss function with respect to the weight at iteration $t$, $\eta$ is the learning rate, $\delta_j$ the error term of the node in the following layer, and $O_j$ the output of that node. The general formula for calculating $\delta_j$ is
\begin{equation}
\delta_j = \frac{dL}{dz_j} = \frac{dL}{dO_j} \cdot \frac{dO_j}{dz_j}
    \label{eq:loss}
\end{equation}
with the loss function $L$ and $z_j$ the pre-activation value of neuron $j$. The first factor $dL/dO_j$ is $O_j-y_{target}$ for the output neuron and $\sum_i \delta_i w_{ij}$ for hidden neurons. The second factor of Eq.~\ref{eq:loss} is the derivative of the activation function, which is $O_j(1-O_j)$ for the sigmoid activation function. \\
As an optimizer we use AdamW \cite{loshchilov_decoupled_2017}, it is a variant of the Adam optimizer \cite{kingma_adam_2014} with an implemented weight decay $\lambda$ which, in comparison to Adam reduces overfitting and improves generalization \cite{zhou_towards_2024} through L2 regularization. As Fig.~\ref{fig: optimizers} shows, larger weights lead to a more aggressive weight update. The weight for iteration $t+1$ without a weight decay is calculated as:
\begin{equation}
    w_{t+1} = w_t - \frac{\eta}{\sqrt{\hat{v}}+\epsilon}\hat{m_t}.
\end{equation}
Where $\hat{m_t}$ is the bias-corrected first moment estimate of $m_t$ which is defined as:
\begin{equation}
    m_t = \beta_1 m_{t-1} + (1 - \beta_1) \frac{dL}{dw_t}.
\end{equation}
$\hat{v}$ is a term to reduce the influence of large weights, $\beta_1$ is a coefficient for computing the first moment estimate and is usually set to 0.9, $\epsilon$ avoids the division by zero. 

For loss calculation, Huber Loss \cite{huber_robust_1964} is used, which combines the advantages of L1Loss and MSELoss by applying the L1Loss for large errors (outliers) and MSELoss for small errors. The threshold is defined by a delta parameter which is set to 1.0 in our case.
\subsection{Hyperparameter Tuning}
\label{ssec:hyper}
Hyperparameters are parameters which define the architecture and the training process of a network, e.g. the number of hidden layers. They are not updated during training. Since they have a huge impact on the GNN performance, it is often beneficial to tune them. There exists a multitude of techniques such as grid search, random search, Bayesian optimization, genetic algorithms, etc. to improve hyperparameters.
\begin{table}[ht]
\centering
\caption{Optimized hyperparameters.}
\begin{tabular}{lc}
    \textbf{Hyperparameter} & \textbf{Value} \\ 
    \midrule
    Embedding size for node features& 18 \\
    Number of convolutional layers&1\\
    Hidden feature size&18\\
    Learning rate&0.003\\
    Weight decay&0.0005\\
    \bottomrule
\end{tabular}
\label{tab:hyperparameters}
\end{table}

A grid search was performed to optimize the hyperparameters of the GNN model (Tab. \ref{tab:hyperparameters}). Batch-size optimization was done manually. Bajaj et al. \cite{bajaj_graph_2024} found that mini-batches achieve faster convergence and higher accuracy compared to the use of the whole dataset.
Hu et al. \cite{hu_batch_2020} propose a batch size of $n/\overline{d}$, where n is the total number of nodes in the graph and the average degree of a graph $\overline{d}$ is calculated by dividing the total number of edges in the graph by the total number of nodes.
Smaller batch sizes lead to more computational costs and therefore a longer training time. On the other hand, using smaller batch sizes can help mitigate memory issues, especially when dealing with large graphs \cite{broadwater_graph_2025}. Also, the update of the gradient is faster: e.g. having 1000 training samples and a batch size of 100, we will update the biases and weights 10 times per epoch. If we used all samples during propagation, there would only be one update per epoch. For a smaller batch size, the gradient estimate will be noisier which can cause the loss to fluctuate more during training. Yet, the noisiness of gradient estimates in small batch sizes can help the optimization process bypass local optima and potentially converge to a global optimum \cite{neelakantan_adding_2015}.\\
In the presented GNN model, the learning rate is set to \SI{9e-3}{}, the batch size to 100, the number of epochs to 150, and the weight decay for the AdamW optimizer to \SI{5e-4}{}.

\section{Uncertainty quantification with Bayesian GNN}
\label{sec: uq with bnn}
Since we aim to predict the coercive field for microstructures that have not been simulated and therefore have no known label, it is important to assess how confident the model is in its predictions. Therefore, uncertainty quantification plays a crucial role. 
To calculate the uncertainty in GNNs, there exists a variety of techniques \cite{gawlikowski_survey_2023}. We used a Bayesian approximation method, namely Monte-Carlo (MC) dropout, to quantify uncertainty in our GNN. MC dropout implies that we use active dropout layers during prediction to capture epistemic uncertainty. Here, epistemic uncertainty refers to uncertainty in the model parameters due to limited data, and can be reduced with more training data. In contrast, aleatoric uncertainty arises from inherent noise in the data itself and cannot be reduced even with more data. The Gaussian negative log likelihood loss \cite{nix_estimating_1994} function allows us to additionally capture the aleatoric uncertainty \cite{kendall_what_2017}.
We assume that the model's input $x$ 
has a varying noise level for different input values. And we assume the target $y$ to have normally distributed errors. Then we will get for the target probability distribution

\begin{align}
\label{eq: gaussian distr}
    P(y_i|x_i,\text{GNN}) = \\ \nonumber \frac{1}{\sqrt{2\pi \sigma^2(x_i)}}\exp \left (\frac{-(y_i-\mu(x_i))^2}{2\sigma²(x_i)} \right).
\end{align}
Eq. \ref{eq: gaussian distr} represents the predictive distribution of the output $y_i$, given the input $x_i$ and the trained GNN model. $\mu$ is the mean of the target and $\sigma^2 = \text{Var}[y_i | x_i]$ is the variance of the distribution. 
If we take the negative of the natural log of Eq.~\ref{eq: gaussian distr}, it results in our new loss function:
\begin{align}
\label{eq: gnllloss}
    -\ln  P(y_i|x_i,\text{GNN}) = \\ \frac{1}{2}\ln(2\pi) + \nonumber \frac{1}{2} \ln(\sigma^2(x_i)) + \frac{(y_i-\mu(x_i))^2}{2\sigma²(x_i)}
\end{align}
where the first term can be omitted because it's constant. 
The loss function is encouraging the model to accurately predict $\mu(x_i)$ thus minimizing the third term. However, if the prediction of $\mu(x_i)$ is poor, increasing the variance will reduce the third term again. And the second term prevents the model from predicting an excessively large variance. 

The model's predicted mean $\mu(x_i)$ of the target approximates the conditional expected value $\mathbb{E}\left[y^*|\mathbf{x}^*,\theta\right]$. Then, the variance $\mathbb{V}$ of the total uncertainty in the prediction of a GNN is given by summing the variances of the epistemic and aleatoric uncertainties:
\begin{equation}
\label{eq: gaussian distr log}
    \mathbb{V}(y^*|\mathbf{x}^*, \mathcal{D}) = \\
    \underbrace{\mathbb{E}_{\theta|\mathcal{D}}\left[\mathbb{V}\left(y^*|\mathbf{x}^*,\theta\right]\right)}_{\text{Aleatoric}} + \\
    \underbrace{\mathbb{V}_{\theta|\mathcal{D}}(\mathbb{E}\left[y^*|\mathbf{x}^*,\theta\right])}_{\text{Epistemic}}
    \end{equation}
This follows from the law of total variance, a general formula for variance decomposition.
Here $y^*$ is a label, predicted from an instance of features $\mathbf{x^*}$. $\theta$ are the model parameters and $\mathcal{D}$ is the dataset on which the model is trained.
In our case, aleatoric uncertainty of predicted coercivity comes from the features that non-uniquely describe the total polycrystal structure. Therefore, the same set of input features may result in a different computed graph label. Epistemic uncertainty, on the other hand, results from the model's imperfection in predicting target values, often due to limited training data or an inadequate model architecture \cite{kiureghian_aleatory_2009}. 

For uncertainty prediction, the model was trained using the Gaussian negative log likelihood (Eq.~\ref{eq: gnllloss}) as the loss function, and then predictions were made with active dropout layers in the post convolution layer \cite{ryu_bayesian_2019}.
We predicted each point in the test set 50 times. The variance of the predicted mean target values $\mu(x_i)$ represents the epistemic uncertainty, and the mean of the predicted variances $\sigma^2$ represents the aleatoric uncertainty. This means, to get an idea of the model's imperfection, one has to predict multiple times for the same input value.

\begin{figure}
    \centering
	\includegraphics[width=3.4in]{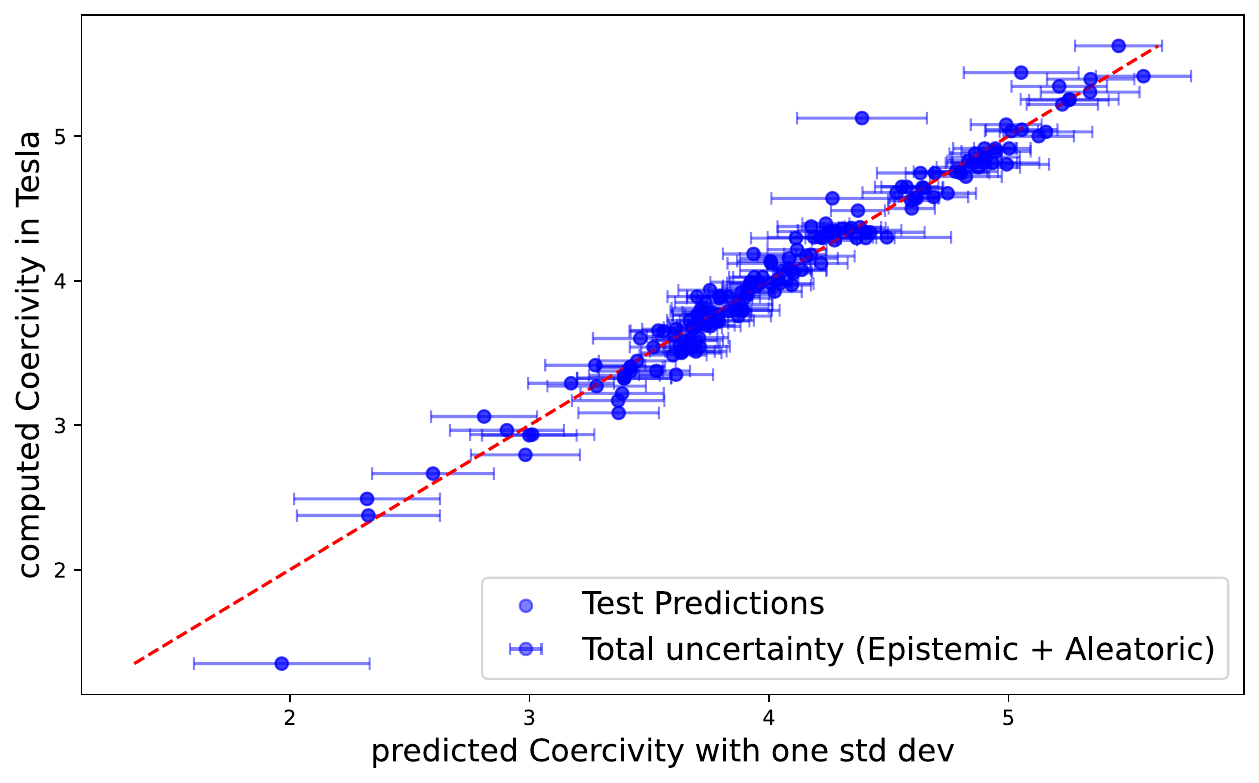}
	\caption{The total predictive uncertainty of the test set. The blue dots are the predicted mean and the lines represent one standard deviation.}
	\label{fig:total uncertainty}
\end{figure}

\begin{figure}
    \centering
	\includegraphics[width=3.4in]{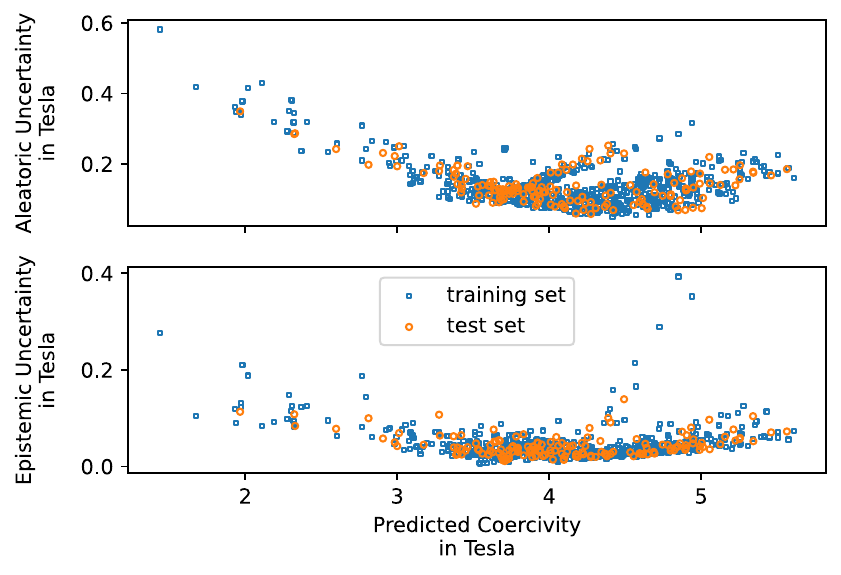}
	\caption{The aleatoric and epistemic uncertainty of the training and test set.}
	\label{fig:aleatoric and epistemic}
\end{figure}

\begin{figure}
    \centering
	\includegraphics[width=3.4in]{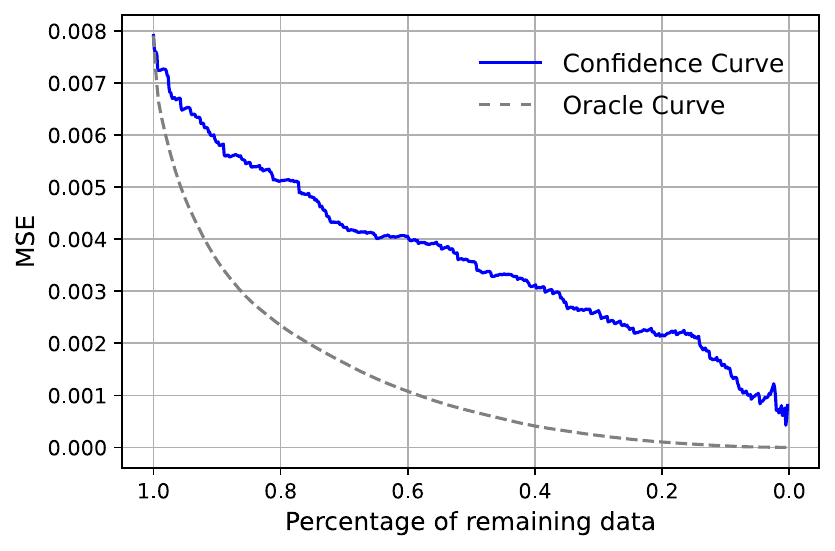}
	\caption{Confidence curve for the training set. The blue line represents the MSE for removing percentages of data, sorted according to uncertainty. The dashed line represents MSE for removing percentages of data, sorted according to true prediction error, also called oracle curve.}
	\label{fig:confidence curve}
\end{figure}

Fig.~\ref{fig:total uncertainty} shows the predicted coercivity against the true computed coercivity (blue dots). The error bars show one standard deviation of the total uncertainty, meaning the model expects that 68\,\% of the true values fall within these bars, assuming a normal distribution of the uncertainty. Fig.~\ref{fig:aleatoric and epistemic} depicts the aleatoric and epistemic uncertainty for the training and test set. It can be seen that, especially for values with lower coercivity, the uncertainty is larger. This is presumably due to the sparser dataset and the origins of lower coercivity, such as larger grains. The uncertainty reaches a minimum at around four Tesla. In general, the aleatoric uncertainty is higher than the epistemic, ranging from 0.6 Tesla to nearly zero Tesla, whereas the epistemic range is from 0.4 to zero Tesla. A diagnostic tool for UQ validation \cite{pernot_confidence_2022} is a confidence curve (CC). To construct a confidence curve (CC), the dataset is first sorted based on the uncertainty estimates of each sample. The sample with the highest uncertainty is then removed, and the mean squared error (MSE) of the remaining predictions is computed. This process is repeated iteratively, progressively excluding the most uncertain samples. The obtained curve should be continuously decreasing to reveal a proper association between the prediction errors and the UQ. The blue line in Fig.~\ref{fig:confidence curve} shows a CC for the training set of the GNN. It can be seen that the curve is steadily decreasing; therefore, showing that the predicted uncertainties can be associated with the prediction errors. This technique could be used to compare the meaningfulness between other models' and architectures' uncertainty estimates.

\section{Results}
\label{sec: results}
We trained the developed GNN on a dataset of over 1000 hard magnetic microstructures. The results for the coercivity prediction are presented in subsection~\ref{ssec:coercivity_prediction}. Furthermore, we derived the maximum energy product from the hysteresis curves of the dataset and adjusted our GNN to predict it in subsection~\ref{ssec:bhmax_prediction}. Finally, in subsection~\ref{ssec:ood}, we examine the out-of-distribution capabilities of the GNN trained with coercivity as the target.
\subsection{Coercivity prediction}
\label{ssec:coercivity_prediction}
\begin{figure}
    \centering
	\includegraphics[width=3.4in]{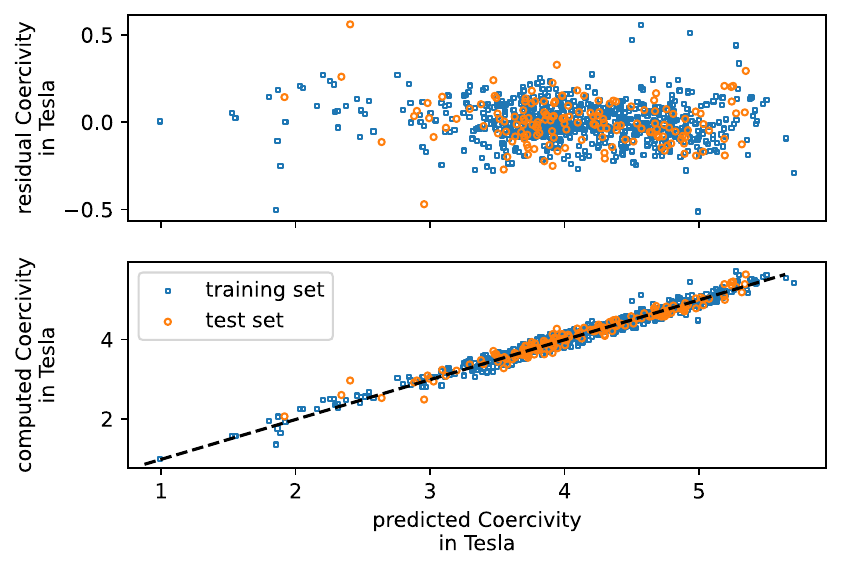}
	\caption{Difference between prediction and computation of the coercivity, and predicted versus computed coercivity. The dashed line represents zero-error for prediction. $\text{R}^2 = 96\,\%$ for the test set.}
	\label{fig:true_vs_predicted_hc}
\end{figure}

Fig.~\ref{fig:true_vs_predicted_hc} shows the difference between prediction and computation of the coercivity in the lower plot for the training and test set. Points that are on the diagonal dashed line are predicted without an error. Most microstructures have a coercivity between 3.5 and 5.5 Tesla; however, the lowest coercivity of a microstructure is approximately one, and even that is accurately predicted. The coefficient of determination, also known as the R² value of the test set, is 96\,\% which underlines the good performance of the GNN for predicting coercivity. The upper plot shows the residual of the coercivity (computed minus predicted). The residual is varying from -0.5 Tesla to 0.5 Tesla, but no clear trend is identifiable. 

\subsection{GNN Recycling -- Maximum energy product prediction}
\label{ssec:bhmax_prediction}
\begin{figure}
    \centering
	\includegraphics[width=3.4in]{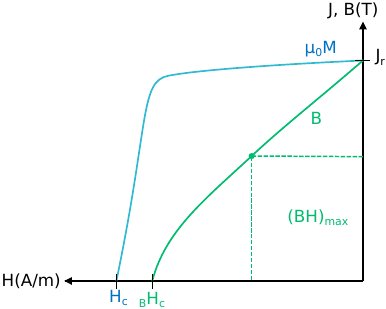}
    \caption{Hysteresis curve $\mu_0\mathrm{M}$ and desheared hysteresis curve $\mathrm{B}$. $\mathrm{J_r}$ is the remanent magnetic polarization, $\mathrm{J}$ and $\mathrm{B(T)}$ correspond to the magnetic polarization and magnetic flux density, respectively, $\mathrm{H(A/m)}$ to the internal magnetic field, and $\mathrm{H_c}$ and $_\mathrm{B}\mathrm{H}_{\mathrm{c}}$ are the coercivities. Image used with kind permission from the MaMMoS Project~\cite{MaMMoS_Ontology}.}
	\label{fig:bhmax}
\end{figure}

\begin{figure}
    \centering
	\includegraphics[width=3.4in]{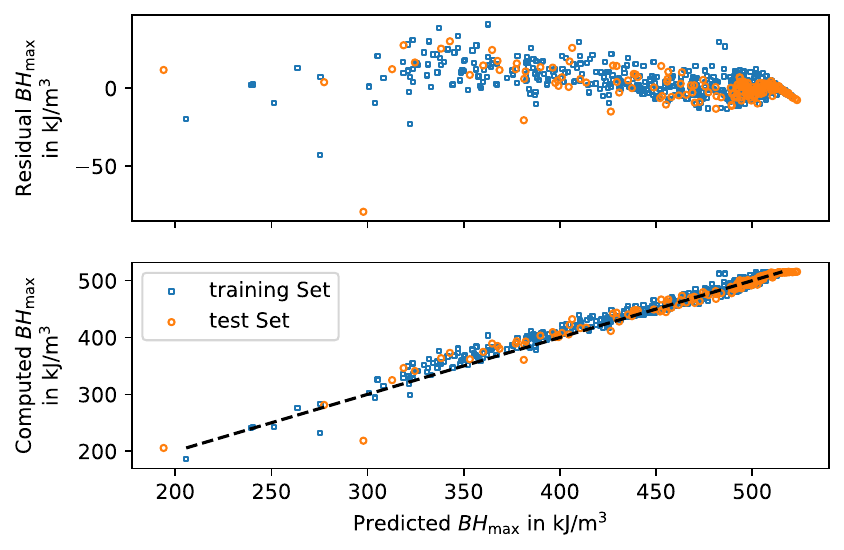}
	\caption{Difference between prediction and computation of the maximum energy product $BH_\mathrm{max}$, and predicted versus computed $BH_\mathrm{max}$. The dashed line represents zero-error for prediction. R² = 97\,\% for the test set.}
	\label{fig:true_vs_predicted_bhmax}
\end{figure}

Given that the architecture of the GNN was already established, we changed the graph's label to predict the maximum energy product, $BH_\mathrm{max}$, thereby changing the existing framework to another target property. $BH_\mathrm{max}$ is defined as the area of the largest rectangle that can be drawn into the second quadrant of the desheared BH-loop (see Fig. \ref{fig:bhmax}). Deshearing is performed with a demagnetizing factor of
$N=1/3$, corresponding to that of a cube. Consequently, the externally applied magnetic field $H$ is calculated as $H = H_{tot}-1/3\cdot M$ where $H_{tot}$ is the total magnetic field strength. The magnetic flux density is calculated as $B=\mu_0(H+M)$. $BH_\mathrm{max}$ represents the energy density and therefore the technically usable energy in the magnet.
The dataset used for $BH_\mathrm{max}$ prediction consists of 1206 microstructures, which is slightly larger than the one used for coercivity prediction. This difference arises because the maximum energy product prediction was conducted at a later stage when more simulations existed.
The GNN architecture remained the same, however, six additional input features were found to increase the prediction accuracy. \begin{itemize}
    \item The volume, 
    \item the equivalent diameter which is the diameter of a sphere with the same volume as the original grain,
    \item the convexity which is the ratio of the volume of the grain to the volume of the convex hull of the grain, 
    \item the minimal dihedral angle, 
    \item the maximal dihedral angle,
    \item and the number of neighbors.
\end{itemize} 
The lower Fig.~\ref{fig:true_vs_predicted_bhmax} shows the predicted versus computed values of the maximum energy product. With an $R^2$ coefficient of 97\,\% the prediction is very good. The upper Fig.~\ref{fig:true_vs_predicted_bhmax} depicts the residual, where one can see that for lower $BH_\mathrm{max}$ values, the accuracy of the prediction decreases slightly.

\subsection{Out-of-distribution Generalization}
\label{ssec:ood}
To enable the prediction of larger unseen microstructures, and to account for potential uncontrolled and unknown shifts in the test set feature distribution, known as Out-of-distribution (OOD) shifts, which pose a risk of reduced model prediction accuracy \cite{fan_generalizing_2024}, we evaluated the model's OOD generalization capabilities. Improving the generalization capabilities for structures with larger edge lengths is particularly important for us, as micromagnetic simulations of these structures are computationally expensive.

There exists a variety of techniques to improve the OOD generalization capabilities of GNN \cite{li_out--distribution_2022}. As part of feature engineering, we found through iterative testing that better OOD generalization can be achieved by scaling edge-length-dependent features with $\log_{1.05}(1+\mathrm{edge\ length/100})$ rather than multiplying them directly by the edge length.  For instance, features such as area —- provided by Neper in a normalized form -— are dependent on the edge length. The scaling was applied to the three coordinates, the volume -- with an exponent of three --, the equivalent diameter, and the area with an exponent of two. This transformation accounts for the observed logarithmic relationship between coercivity and system size (see Section~\ref{sec:size}).

\begin{figure}
    \centering
	\includegraphics[width=3.4in]{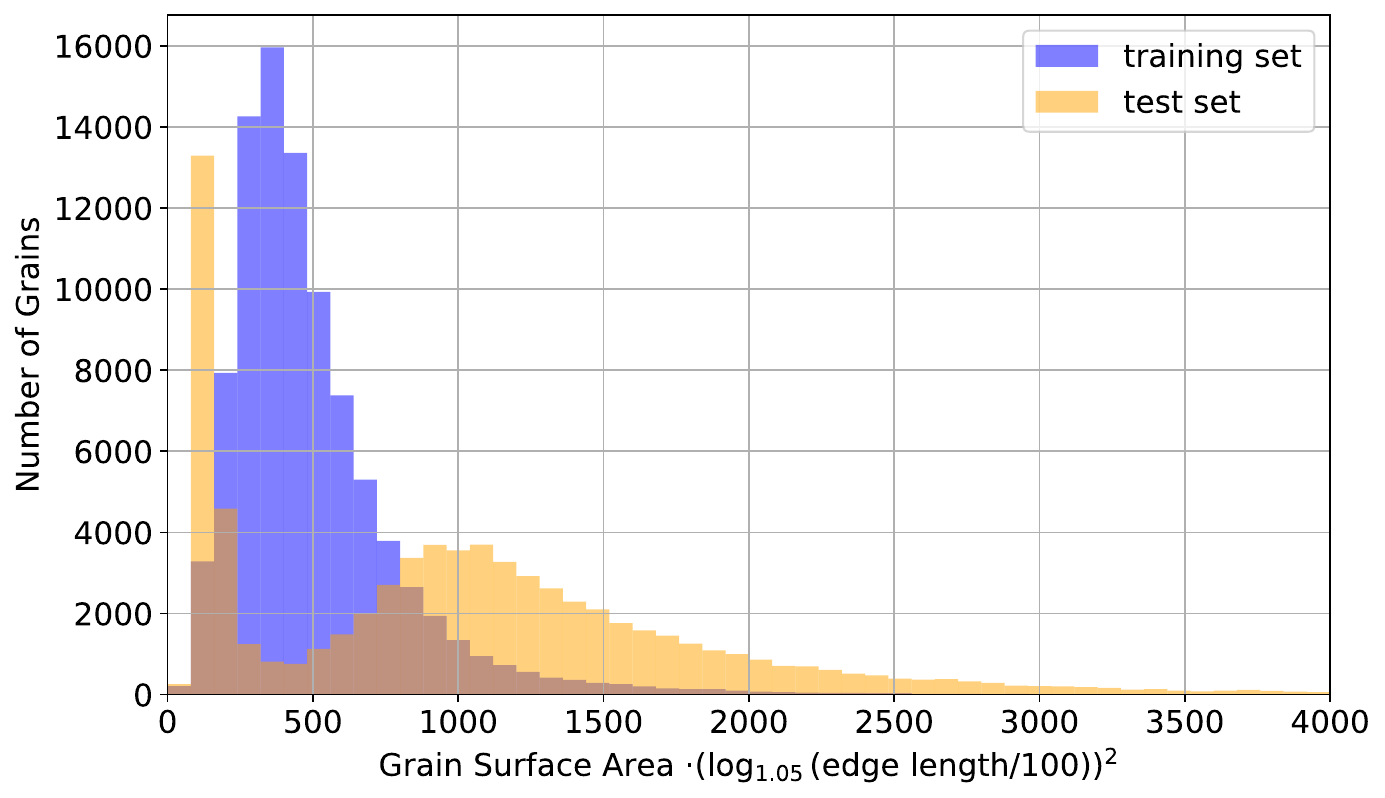}
	\caption{Distribution of the log-scaled grain surface areas for the training and test set. Compared to the training set, the test set has a distribution shift towards larger log-scaled grain surface areas.}
	\label{fig: grain area}
\end{figure}

The simulated dataset consists of 1,066 cubic polycrystalline structures, including 12 larger structures added to increase variability. It was split into training, validation, and test sets containing 648 (61\,\%), 278 (26\,\%), and 140 (13\,\%) structures, respectively. The training and validation sets contain structures that have an edge length below 1.5\,µm. The test set contains all structures that are larger. Fig.~\ref{fig: grain area} shows the distribution of the log-scaled grain surface areas for the training and test set. A distributional shift towards larger grain surface areas can be observed in the test set compared to the training set.

\begin{figure}
    \centering
	\includegraphics[width=3.4in]{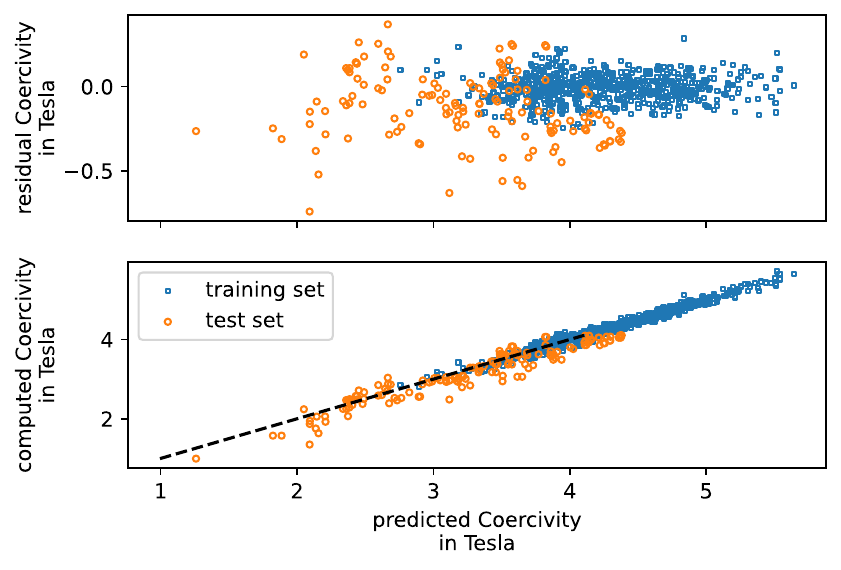}
	\caption{The size generalization test was conducted by training and validating the model on structures with an edge length lower than 1500\,nm, while testing it on structures with an edge length larger than 1500\,nm. On the test dataset, the best-performing model achieved an R² of 88\,\%.}
	\label{fig: resplot_generalization}
\end{figure}
After extensive training, the resulting $\text{R}^2$ coefficient for the best-performing model on the test set is 88\,\% (see Fig.~\ref{fig: resplot_generalization}), which is a good result and proves the OOD generalization capabilities of the GNN with the log-scaled features.

\section{Influence of the system Size on Coercivity}
\label{sec:size}
To investigate the influence of larger polycrystalline structures on coercivity, cubic structures with a varying edge length, but otherwise constant parameters, were simulated. The dataset contains 291 cubic polycrystalline structures with an edge length from 200\,nm to 3000\,nm, an aspect ratio of one and a grain boundary thickness of 5\,nm. A structure with an edge length of 200\,nm contains one grain; consequently, a structure with an edge length of 3000\,nm contains 3375 grains. The easy axis is defined for each grain using a random uniform distribution within a pre-defined maximal cone angle of 15\degree.

\begin{figure}
    \centering
	\includegraphics[width=3.4in]{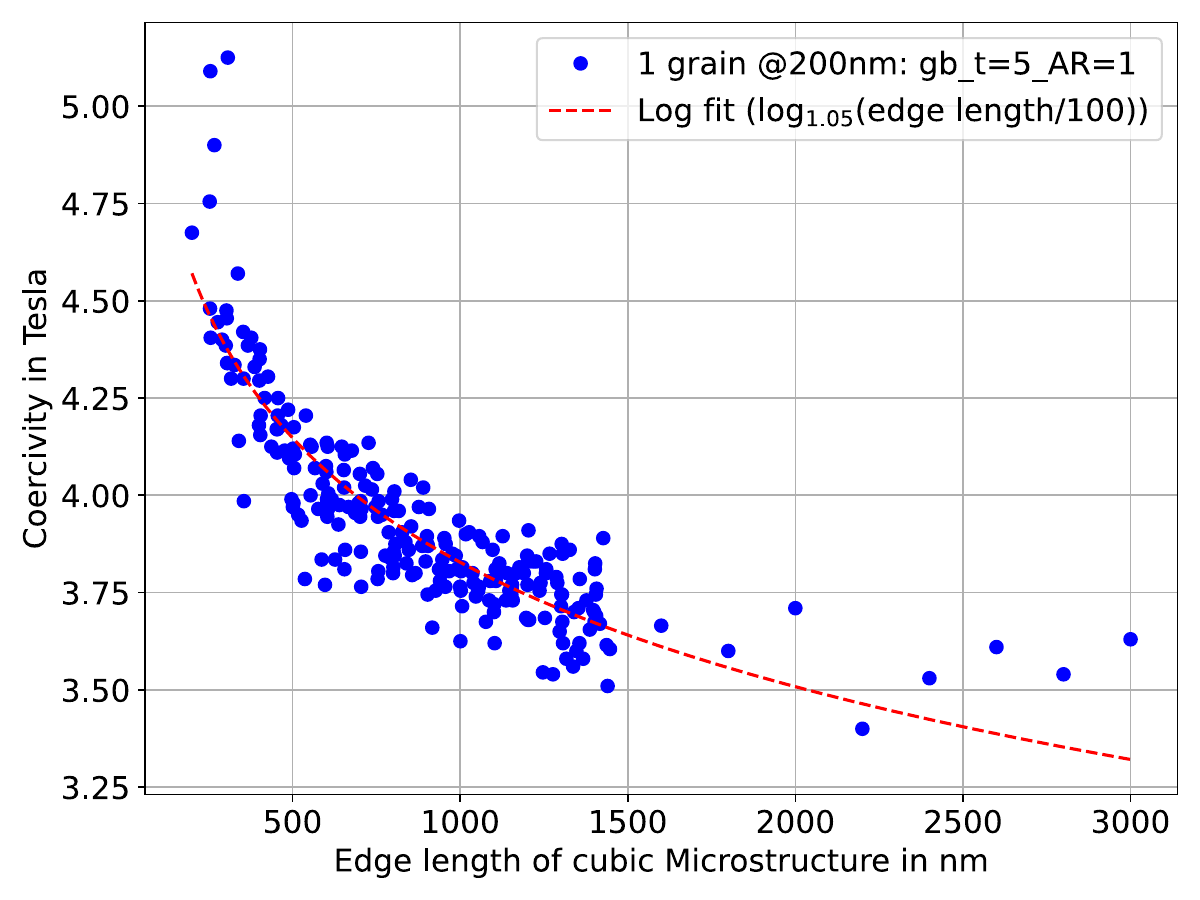}
	\caption{Coercivity as a function of the cubes' edge length. The fitting function is $H_c = -2.251\cdot 10^{-2}\cdot \log_{1.05}(x/100)+4.89$, where $x$ is the edge length in nm.}
	\label{fig:size_coercivity}
\end{figure}

The simulated coercivity decreases when the system size increases (Fig.~\ref{fig:size_coercivity}). The evaluated fitting function $H_c = -2.251\cdot 10^{-2}\cdot \log_{1.05}(x/100)+4.89$, where $x$ is the edge length in nm, demonstrates good agreement with the simulated data within the investigated range.

\begin{figure}[htb]
    \centering
    \begin{subfigure}[t]{0.3\textwidth}
        \centering
        \includegraphics[width=\textwidth]{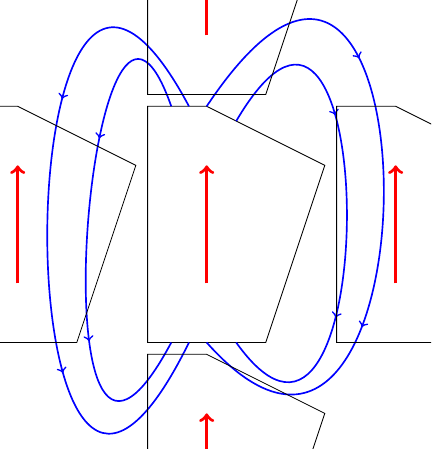}
        \caption{The center grain is not switched.}
        \label{fig:left}
    \end{subfigure}
    \hspace{0.03\textwidth}
    \begin{subfigure}[t]{0.3\textwidth}
        \centering
        \includegraphics[width=\textwidth]{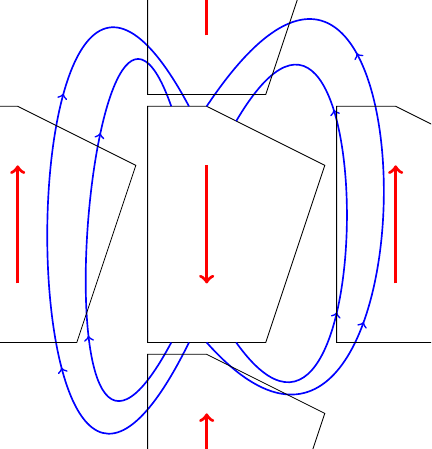}
        \caption{The center grain is switched.}
        \label{fig:right}
    \end{subfigure}
    \caption{Blue lines depict the stray field and red arrows the grains' magnetization. In (a) the stray field is opposed to the magnetization of the horizontal neighbors' magnetization, weakening them. In (b) where the center grain is switched, the stray field is adding to their magnetization and therefore strengthening them whilst weakening the vertical neighbors. This causes the cascade-like switching effect.}
    \label{fig:cascade}
\end{figure}

The main reason for the decreased coercivity is, as shown by \cite{bance_grain-size_2014}, that the demagnetizing perpendicular stray field component close to the edges of a grain increases with grain size. This results -- due to the changed angle of the total field -- in a reduction of the switching field of the grains near the microstructure's edge. 
Another reason for the decreased coercivity is the cascade-like effect (see Fig.~\ref{fig:cascade}) if one grain switches, the ones below and above it are more likely to switch as well, because the inverted magnetostatic field of the switched grain will counteract their magnetization even further. The grains on the side will be stabilized, because the magnetostatic field of the switched grain will support their magnetization. If a critical amount of grains is switched, the destabilizing effect will become larger. Especially for larger grains, where the magnetostatic field has a higher influence due to the size-dependent energy competition between exchange energy and magnetostatic energy, all grains may switch at one specific applied field; this is called single-step switching. This distribution of the coercive field for large grain size \cite{kovacs_physics-informed_2023} is reflected in the higher uncertainty for low coercive field values discussed in Sec. \ref{sec: uq with bnn}. The weakest grain is initializing the single-step switching. Due to random grain arrangement in each microstructure and random easy axis assignment, the local switching field of the weakest grain is varying.\\

\section{Conclusion}
\label{sec: conclusion}
In this work, GNNs are found to be a viable tool to predict coercivity and the maximum energy product. We generated a diverse dataset of micromagnetic structures. The simulations, to obtain this dataset, were done using our reduced order model.

Data preprocessing is essential to ensure that the GNN can learn meaningful patterns. Correlation analysis, feature selection, and feature scaling help the model improve performance. The use of permutation feature importance helped to understand the features' influence on the model performance and highlighted the connection to the underlying physics of magnetization reversal.

By embedding node features into a higher-dimensional space and incorporating both node and edge information through a customized convolutional layer, the model effectively captures the complex interactions within microstructures. The use of dropout helps to prevent overfitting. The GNN architecture was kept shallow, with a single convolutional layer, to preserve the distinctiveness of node features. This ensures that the neighborhood information is effectively captured. AdamW was used as the optimizer in combination with the Huber loss, exploiting their respective strengths in stable convergence and robustness to outliers. This architecture, in combination with hyperparameter tuning, enables the GNN to learn meaningful representations for accurate coercivity prediction.

To ensure reliable predictions for unseen microstructures, uncertainty quantification was integrated into the GNN using Monte Carlo dropout and Gaussian negative log likelihood loss. This approach enabled the model to capture both epistemic and aleatoric uncertainties. The analysis revealed that uncertainty is higher for microstructures with lower coercivity, likely due to data sparsity and structural variability. Aleatoric uncertainty presumably arises from the non-unique representation of features characterizing the polycrystalline structure. A confidence curve further validated the uncertainty estimates, demonstrating their strong correlation with actual prediction errors.

The trained GNN demonstrated strong predictive performance across multiple magnetic properties. For coercivity, the model achieved an $\text{R}^2$ of 96\,\%, accurately capturing a wide range of coercivity values. By reusing the same architecture and extending the feature set, the GNN also successfully predicted the maximum energy product $\text{BH}_{max}$ with an $\text{R}^2$ of 97\,\%, exhibiting its adaptability to different target properties. Furthermore, the model demonstrated good out-of-distribution generalization, achieving an  $\text{R}^2$ of 88\,\%, on significantly larger microstructures. This was achieved by feature scaling, inspired by simulations that demonstrate the coercivity decrease with increasing system size, following a logarithmic trend. These findings underscore the importance of incorporating size effects into predictive models. These results highlight the model’s versatility, accuracy, and potential for practical application in magnetic materials research.

\section*{Acknowledgement}
This research was funded in whole or in part by the Austrian Science Fund (FWF) [10.55776/I6159]. Furthermore, the financial support by the Austrian Federal Ministry of Economy, Energy and Tourism, the National Foundation for Research, Technology and Development, and the Christian Doppler Research Association is gratefully acknowledged. For open access purposes, the author has applied a CC BY public copyright license to any author-accepted manuscript version arising from this submission.

\section*{Data Availability}
The data and code used in this study are openly available on GitHub at: \url{https://github.com/heisammoustafa/Micromagnetic_GNN}

\bibliographystyle{elsarticle-num}
\bibliography{refs}

\begin{thebibliography}{10}
\expandafter\ifx\csname url\endcsname\relax
  \def\url#1{\texttt{#1}}\fi
\expandafter\ifx\csname urlprefix\endcsname\relax\def\urlprefix{URL }\fi
\expandafter\ifx\csname href\endcsname\relax
  \def\href#1#2{#2} \def\path#1{#1}\fi

\bibitem{skomski_magnetic_2016}
R.~Skomski, J.~Coey, Magnetic anisotropy — {How} much is enough for a permanent magnet?, Scripta Materialia 112 (2016) 3--8.
\newblock \href {https://doi.org/10.1016/j.scriptamat.2015.09.021} {\path{doi:10.1016/j.scriptamat.2015.09.021}}.

\bibitem{poudyal_advances_2013}
N.~Poudyal, J.~Ping~Liu, Advances in nanostructured permanent magnets research, Journal of Physics D: Applied Physics 46~(4) (2013) 043001.
\newblock \href {https://doi.org/10.1088/0022-3727/46/4/043001} {\path{doi:10.1088/0022-3727/46/4/043001}}.

\bibitem{sanchez-gonzalez_a_godwin_j_pfaff_t_ying_r_leskovec_j_battaglia_pw_learning_2020}
A.~Sanchez-Gonzalez, J.~Godwin, T.~Pfaff, R.~Ying, J.~Leskovec, P.~Battaglia, Learning to simulate complex physics with graph networks, in: Proceedings of the 37th International Conference on Machine Learning, Vol. 119, 2020, pp. 8459--8468.

\bibitem{dai_graph_2023}
M.~Dai, M.~F. Demirel, X.~Liu, Y.~Liang, J.-M. Hu, \href{https://linkinghub.elsevier.com/retrieve/pii/S092702562300455X}{Graph neural network for predicting the effective properties of polycrystalline materials: {A} comprehensive analysis}, Computational Materials Science 230 (2023) 112461.
\newblock \href {https://doi.org/10.1016/j.commatsci.2023.112461} {\path{doi:10.1016/j.commatsci.2023.112461}}.
\newline\urlprefix\url{https://linkinghub.elsevier.com/retrieve/pii/S092702562300455X}

\bibitem{qin_grainnn_2023}
Y.~Qin, S.~DeWitt, B.~Radhakrishnan, G.~Biros, \href{https://linkinghub.elsevier.com/retrieve/pii/S0927025622006383}{{GrainNN}: {A} neighbor-aware long short-term memory network for predicting microstructure evolution during polycrystalline grain formation}, Computational Materials Science 218 (2023) 111927.
\newblock \href {https://doi.org/10.1016/j.commatsci.2022.111927} {\path{doi:10.1016/j.commatsci.2022.111927}}.
\newline\urlprefix\url{https://linkinghub.elsevier.com/retrieve/pii/S0927025622006383}

\bibitem{hestroffer_graph_2023}
J.~M. Hestroffer, M.-A. Charpagne, M.~I. Latypov, I.~J. Beyerlein, \href{https://linkinghub.elsevier.com/retrieve/pii/S092702562200605X}{Graph neural networks for efficient learning of mechanical properties of polycrystals}, Computational Materials Science 217 (2023) 111894.
\newblock \href {https://doi.org/10.1016/j.commatsci.2022.111894} {\path{doi:10.1016/j.commatsci.2022.111894}}.
\newline\urlprefix\url{https://linkinghub.elsevier.com/retrieve/pii/S092702562200605X}

\bibitem{hamilton_inductive_2017}
W.~Hamilton, Z.~Ying, J.~Leskovec, \href{https://proceedings.neurips.cc/paper_files/paper/2017/file/5dd9db5e033da9c6fb5ba83c7a7ebea9-Paper.pdf}{Inductive {Representation} {Learning} on {Large} {Graphs}}, in: I.~Guyon, U.~V. Luxburg, S.~Bengio, H.~Wallach, R.~Fergus, S.~Vishwanathan, R.~Garnett (Eds.), Advances in {Neural} {Information} {Processing} {Systems}, Vol.~30, Curran Associates, Inc., 2017.
\newline\urlprefix\url{https://proceedings.neurips.cc/paper_files/paper/2017/file/5dd9db5e033da9c6fb5ba83c7a7ebea9-Paper.pdf}

\bibitem{NEURIPS2019_9015}
A.~Paszke, S.~Gross, F.~Massa, A.~Lerer, J.~Bradbury, G.~Chanan, T.~Killeen, Z.~Lin, N.~Gimelshein, L.~Antiga, A.~Desmaison, A.~Kopf, E.~Yang, Z.~DeVito, M.~Raison, A.~Tejani, S.~Chilamkurthy, B.~Steiner, L.~Fang, J.~Bai, S.~Chintala, Pytorch: An imperative style, high-performance deep learning library, in: Advances in Neural Information Processing Systems 32, Curran Associates, Inc., 2019, pp. 8024--8035.

\bibitem{fey_fast_2019}
M.~Fey, J.~E. Lenssen, \href{https://arxiv.org/abs/1903.02428}{Fast {Graph} {Representation} {Learning} with {PyTorch} {Geometric}}, version Number: 3 (2019).
\newblock \href {https://doi.org/10.48550/ARXIV.1903.02428} {\path{doi:10.48550/ARXIV.1903.02428}}.
\newline\urlprefix\url{https://arxiv.org/abs/1903.02428}

\bibitem{quey_large-scale_2011}
R.~Quey, P.~Dawson, F.~Barbe, Large-scale {3D} random polycrystals for the finite element method: {Generation}, meshing and remeshing, Computer Methods in Applied Mechanics and Engineering 200~(17-20) (2011) 1729--1745.
\newblock \href {https://doi.org/10.1016/j.cma.2011.01.002} {\path{doi:10.1016/j.cma.2011.01.002}}.

\bibitem{salome}
{Salom{\'{e}}}, {The Open Source Integration Platform for Numerical Simulation}, \url{www.salome-platform.org} (Accessed: April 2025) (2025).

\bibitem{moustafa_reduced_2024}
H.~Moustafa, A.~Kovacs, J.~Fischbacher, M.~Gusenbauer, Q.~Ali, L.~Breth, Y.~Hong, W.~Rigaut, T.~Devillers, N.~M. Dempsey, T.~Schrefl, H.~Oezelt, \href{https://pubs.aip.org/adv/article/14/2/025001/3261431/Reduced-order-model-for-hard-magnetic-films}{Reduced order model for hard magnetic films}, AIP Advances 14~(2) (2024) 025001.
\newblock \href {https://doi.org/10.1063/9.0000816} {\path{doi:10.1063/9.0000816}}.
\newline\urlprefix\url{https://pubs.aip.org/adv/article/14/2/025001/3261431/Reduced-order-model-for-hard-magnetic-films}

\bibitem{singh_new_2001}
B.~Singh, D.~Guptasarma, \href{https://library.seg.org/doi/10.1190/1.1444942}{New method for fast computation of gravity and magnetic anomalies from arbitrary polyhedra}, GEOPHYSICS 66~(2) (2001) 521--526.
\newblock \href {https://doi.org/10.1190/1.1444942} {\path{doi:10.1190/1.1444942}}.
\newline\urlprefix\url{https://library.seg.org/doi/10.1190/1.1444942}

\bibitem{mikhalev_iterative_2016}
A.~Y. Mikhalev, I.~V. Oseledets, \href{https://onlinelibrary.wiley.com/doi/10.1002/nla.2021}{Iterative representing set selection for nested cross approximation}, Numerical Linear Algebra with Applications 23~(2) (2016) 230--248.
\newblock \href {https://doi.org/10.1002/nla.2021} {\path{doi:10.1002/nla.2021}}.
\newline\urlprefix\url{https://onlinelibrary.wiley.com/doi/10.1002/nla.2021}

\bibitem{marsaglia_choosing_1972}
G.~Marsaglia, \href{http://projecteuclid.org/euclid.aoms/1177692644}{Choosing a {Point} from the {Surface} of a {Sphere}}, The Annals of Mathematical Statistics 43~(2) (1972) 645--646.
\newblock \href {https://doi.org/10.1214/aoms/1177692644} {\path{doi:10.1214/aoms/1177692644}}.
\newline\urlprefix\url{http://projecteuclid.org/euclid.aoms/1177692644}

\bibitem{geron_hands-machine_2019}
A.~Géron, Hands-on machine learning with {Scikit}-{Learn}, {Keras}, and {TensorFlow}: concepts, tools, and techniques to build intelligent systems, second edition Edition, O'Reilly Media, Inc, Beijing [China] ; Sebastopol, CA, 2019.

\bibitem{scikit-learn}
F.~Pedregosa, G.~Varoquaux, A.~Gramfort, V.~Michel, B.~Thirion, O.~Grisel, M.~Blondel, P.~Prettenhofer, R.~Weiss, V.~Dubourg, J.~Vanderplas, A.~Passos, D.~Cournapeau, M.~Brucher, M.~Perrot, E.~Duchesnay, Scikit-learn: Machine learning in {P}ython, Journal of Machine Learning Research 12 (2011) 2825--2830.

\bibitem{noauthor_42_nodate}
\href{https://scikit-learn.org/stable/modules/permutation_importance.html}{4.2. {Permutation} feature importance}.
\newline\urlprefix\url{https://scikit-learn.org/stable/modules/permutation_importance.html}

\bibitem{hinton_improving_2012}
G.~E. Hinton, N.~Srivastava, A.~Krizhevsky, I.~Sutskever, R.~R. Salakhutdinov, \href{http://arxiv.org/abs/1207.0580}{Improving neural networks by preventing co-adaptation of feature detectors}, arXiv:1207.0580 [cs] (Jul. 2012).
\newblock \href {https://doi.org/10.48550/arXiv.1207.0580} {\path{doi:10.48550/arXiv.1207.0580}}.
\newline\urlprefix\url{http://arxiv.org/abs/1207.0580}

\bibitem{lai_analysis_2017}
S.~Park, N.~Kwak, \href{http://link.springer.com/10.1007/978-3-319-54184-6_12}{Analysis on the {Dropout} {Effect} in {Convolutional} {Neural} {Networks}}, in: S.-H. Lai, V.~Lepetit, K.~Nishino, Y.~Sato (Eds.), Computer {Vision} – {ACCV} 2016, Vol. 10112, Springer International Publishing, Cham, 2017, pp. 189--204, series Title: Lecture Notes in Computer Science.
\newblock \href {https://doi.org/10.1007/978-3-319-54184-6_12} {\path{doi:10.1007/978-3-319-54184-6_12}}.
\newline\urlprefix\url{http://link.springer.com/10.1007/978-3-319-54184-6_12}

\bibitem{loshchilov_decoupled_2017}
I.~Loshchilov, F.~Hutter, \href{https://arxiv.org/abs/1711.05101}{Decoupled {Weight} {Decay} {Regularization}}, version Number: 3 (2017).
\newblock \href {https://doi.org/10.48550/ARXIV.1711.05101} {\path{doi:10.48550/ARXIV.1711.05101}}.
\newline\urlprefix\url{https://arxiv.org/abs/1711.05101}

\bibitem{kingma_adam_2014}
D.~P. Kingma, J.~Ba, \href{https://arxiv.org/abs/1412.6980}{Adam: {A} {Method} for {Stochastic} {Optimization}}, version Number: 9 (2014).
\newblock \href {https://doi.org/10.48550/ARXIV.1412.6980} {\path{doi:10.48550/ARXIV.1412.6980}}.
\newline\urlprefix\url{https://arxiv.org/abs/1412.6980}

\bibitem{zhou_towards_2024}
P.~Zhou, X.~Xie, Z.~Lin, S.~Yan, \href{https://ieeexplore.ieee.org/document/10480574/}{Towards {Understanding} {Convergence} and {Generalization} of {AdamW}}, IEEE Transactions on Pattern Analysis and Machine Intelligence 46~(9) (2024) 6486--6493.
\newblock \href {https://doi.org/10.1109/TPAMI.2024.3382294} {\path{doi:10.1109/TPAMI.2024.3382294}}.
\newline\urlprefix\url{https://ieeexplore.ieee.org/document/10480574/}

\bibitem{huber_robust_1964}
P.~J. Huber, \href{http://projecteuclid.org/euclid.aoms/1177703732}{Robust {Estimation} of a {Location} {Parameter}}, The Annals of Mathematical Statistics 35~(1) (1964) 73--101.
\newblock \href {https://doi.org/10.1214/aoms/1177703732} {\path{doi:10.1214/aoms/1177703732}}.
\newline\urlprefix\url{http://projecteuclid.org/euclid.aoms/1177703732}

\bibitem{bajaj_graph_2024}
S.~Bajaj, H.~Son, J.~Liu, H.~Guan, M.~Serafini, \href{https://arxiv.org/abs/2406.00552}{Graph {Neural} {Network} {Training} {Systems}: {A} {Performance} {Comparison} of {Full}-{Graph} and {Mini}-{Batch}}, version Number: 4 (2024).
\newblock \href {https://doi.org/10.48550/ARXIV.2406.00552} {\path{doi:10.48550/ARXIV.2406.00552}}.
\newline\urlprefix\url{https://arxiv.org/abs/2406.00552}

\bibitem{hu_batch_2020}
Y.~Hu, A.~Levi, I.~Kumar, Y.~Zhang, M.~Coates, On batch size selection for stochastic training for graph neural networks, ICLR 2021 Conference (Sep. 2020).

\bibitem{broadwater_graph_2025}
K.~Broadwater, Graph {Neural} {Networks} in {Action}, Manning Publications, US, 2025, oCLC: 1482783825.

\bibitem{neelakantan_adding_2015}
A.~Neelakantan, L.~Vilnis, Q.~V. Le, I.~Sutskever, L.~Kaiser, K.~Kurach, J.~Martens, \href{https://arxiv.org/abs/1511.06807}{Adding {Gradient} {Noise} {Improves} {Learning} for {Very} {Deep} {Networks}}, version Number: 1 (2015).
\newblock \href {https://doi.org/10.48550/ARXIV.1511.06807} {\path{doi:10.48550/ARXIV.1511.06807}}.
\newline\urlprefix\url{https://arxiv.org/abs/1511.06807}

\bibitem{gawlikowski_survey_2023}
J.~Gawlikowski, C.~R.~N. Tassi, M.~Ali, J.~Lee, M.~Humt, J.~Feng, A.~Kruspe, R.~Triebel, P.~Jung, R.~Roscher, M.~Shahzad, W.~Yang, R.~Bamler, X.~X. Zhu, \href{https://link.springer.com/10.1007/s10462-023-10562-9}{A survey of uncertainty in deep neural networks}, Artificial Intelligence Review 56~(S1) (2023) 1513--1589.
\newblock \href {https://doi.org/10.1007/s10462-023-10562-9} {\path{doi:10.1007/s10462-023-10562-9}}.
\newline\urlprefix\url{https://link.springer.com/10.1007/s10462-023-10562-9}

\bibitem{nix_estimating_1994}
D.~Nix, A.~Weigend, \href{http://ieeexplore.ieee.org/document/374138/}{Estimating the mean and variance of the target probability distribution}, in: Proceedings of 1994 {IEEE} {International} {Conference} on {Neural} {Networks} ({ICNN}'94), IEEE, Orlando, FL, USA, 1994, pp. 55--60 vol.1.
\newblock \href {https://doi.org/10.1109/ICNN.1994.374138} {\path{doi:10.1109/ICNN.1994.374138}}.
\newline\urlprefix\url{http://ieeexplore.ieee.org/document/374138/}

\bibitem{kendall_what_2017}
A.~Kendall, Y.~Gal, \href{http://arxiv.org/abs/1703.04977}{What {Uncertainties} {Do} {We} {Need} in {Bayesian} {Deep} {Learning} for {Computer} {Vision}?}, arXiv:1703.04977 [cs] (Oct. 2017).
\newblock \href {https://doi.org/10.48550/arXiv.1703.04977} {\path{doi:10.48550/arXiv.1703.04977}}.
\newline\urlprefix\url{http://arxiv.org/abs/1703.04977}

\bibitem{kiureghian_aleatory_2009}
A.~D. Kiureghian, O.~Ditlevsen, \href{https://linkinghub.elsevier.com/retrieve/pii/S0167473008000556}{Aleatory or epistemic? {Does} it matter?}, Structural Safety 31~(2) (2009) 105--112.
\newblock \href {https://doi.org/10.1016/j.strusafe.2008.06.020} {\path{doi:10.1016/j.strusafe.2008.06.020}}.
\newline\urlprefix\url{https://linkinghub.elsevier.com/retrieve/pii/S0167473008000556}

\bibitem{ryu_bayesian_2019}
S.~Ryu, Y.~Kwon, W.~Y. Kim, \href{https://xlink.rsc.org/?DOI=C9SC01992H}{A {Bayesian} graph convolutional network for reliable prediction of molecular properties with uncertainty quantification}, Chemical Science 10~(36) (2019) 8438--8446.
\newblock \href {https://doi.org/10.1039/C9SC01992H} {\path{doi:10.1039/C9SC01992H}}.
\newline\urlprefix\url{https://xlink.rsc.org/?DOI=C9SC01992H}

\bibitem{pernot_confidence_2022}
P.~Pernot, \href{https://arxiv.org/abs/2206.15272}{Confidence curves for {UQ} validation: probabilistic reference vs. oracle}, version Number: 2 (2022).
\newblock \href {https://doi.org/10.48550/ARXIV.2206.15272} {\path{doi:10.48550/ARXIV.2206.15272}}.
\newline\urlprefix\url{https://arxiv.org/abs/2206.15272}

\bibitem{MaMMoS_Ontology}
M.~Project, Magnetic materials ontology, \url{https://github.com/MaMMoS-project/MagneticMaterialsOntology}, accessed: 2025-05-21 (2024).

\bibitem{fan_generalizing_2024}
S.~Fan, X.~Wang, C.~Shi, P.~Cui, B.~Wang, \href{https://ieeexplore.ieee.org/document/10268633/}{Generalizing {Graph} {Neural} {Networks} on {Out}-of-{Distribution} {Graphs}}, IEEE Transactions on Pattern Analysis and Machine Intelligence 46~(1) (2024) 322--337.
\newblock \href {https://doi.org/10.1109/TPAMI.2023.3321097} {\path{doi:10.1109/TPAMI.2023.3321097}}.
\newline\urlprefix\url{https://ieeexplore.ieee.org/document/10268633/}

\bibitem{li_out--distribution_2022}
H.~Li, X.~Wang, Z.~Zhang, W.~Zhu, \href{https://arxiv.org/abs/2202.07987}{Out-{Of}-{Distribution} {Generalization} on {Graphs}: {A} {Survey}}, version Number: 2 (2022).
\newblock \href {https://doi.org/10.48550/ARXIV.2202.07987} {\path{doi:10.48550/ARXIV.2202.07987}}.
\newline\urlprefix\url{https://arxiv.org/abs/2202.07987}

\bibitem{bance_grain-size_2014}
S.~Bance, B.~Seebacher, T.~Schrefl, L.~Exl, M.~Winklhofer, G.~Hrkac, G.~Zimanyi, T.~Shoji, M.~Yano, N.~Sakuma, M.~Ito, A.~Kato, A.~Manabe, \href{https://pubs.aip.org/jap/article/116/23/233903/168049/Grain-size-dependent-demagnetizing-factors-in}{Grain-size dependent demagnetizing factors in permanent magnets}, Journal of Applied Physics 116~(23) (2014) 233903.
\newblock \href {https://doi.org/10.1063/1.4904854} {\path{doi:10.1063/1.4904854}}.
\newline\urlprefix\url{https://pubs.aip.org/jap/article/116/23/233903/168049/Grain-size-dependent-demagnetizing-factors-in}

\bibitem{kovacs_physics-informed_2023}
A.~Kovacs, J.~Fischbacher, H.~Oezelt, A.~Kornell, Q.~Ali, M.~Gusenbauer, M.~Yano, N.~Sakuma, A.~Kinoshita, T.~Shoji, A.~Kato, Y.~Hong, S.~Grenier, T.~Devillers, N.~M. Dempsey, T.~Fukushima, H.~Akai, N.~Kawashima, T.~Miyake, T.~Schrefl, Physics-informed machine learning combining experiment and simulation for the design of neodymium-iron-boron permanent magnets with reduced critical-elements content, Frontiers in Materials 9 (2023) 1094055.
\newblock \href {https://doi.org/10.3389/fmats.2022.1094055} {\path{doi:10.3389/fmats.2022.1094055}}.

\end{thebibliography}

\end{document}